\documentclass[10pt,twocolumn,showpacs,nofootinbib,aps,prd]{revtex4-1}
\usepackage{CJK}
\usepackage{amsfonts,amsmath,amssymb,mathrsfs}
\usepackage{mathtools,mathptmx,mathenv,array,latexsym}
\usepackage[colorlinks=true,linkcolor=blue,citecolor=blue,urlcolor=blue]{hyperref}
\def\be{\begin{equation}}
\def\ee{\end{equation}}
\def\bea{\begin{eqnarray}}
\def\eea{\end{eqnarray}}
\def\ba{\begin{aligned}}
\def\ea{\end{aligned}}
\def\nn{\nonumber}
\def\cA{\mathcal{A}}

\def\cP{\mathcal{P}}
\def\cV{\mathcal{V}}
\def\cN{\mathcal{N}}
\def\tM{\widetilde{M}}

\def\tph{\widetilde{\psi_h}}
\def\p{\partial}

\def\sta2{\sin^2\theta}

\begin{document}
%\begin{CJK*}{GBK}{gbsk}
%\begin{CJK*}{GBK}{song}

\title{Consistent mass formulas for the four-dimensional dyonic NUT-charged spacetimes}

%\author{Di Wu (ÎâµÏ)}
\author{Di Wu}
\email{wdcwnu@163.com}
% https://orcid.org/0000-0002-2509-6729

%\author{Shuang-Qing Wu (ÎâË«Çå)}
\author{Shuang-Qing Wu}
\email{sqwu@cwnu.edu.cn, Corresponding author}
% https://orcid.org/0000-0001-7936-7195

\affiliation{School of Physics and Astronomy, China West Normal University, Nanchong, Sichuan 637002,
People's Republic of China}

\date{\today}
%\date{Received \today; published -- mm, 2022}

\begin{abstract}
In our previous work [\href{http://dx.doi.org/10.1103/PhysRevD.100.101501}{Phys. Rev. D. \textbf{100},
101501(R) (2019)}], a novel idea that the Newman-Unti-Tamburino (NUT) charge can be thought of as
a thermodynamical multi-hair has been advocated to describe perfectly the thermodynamical character
of the generic four-dimensional Taub-NUT spacetimes. According to this scheme, the Komar mass ($M$),
the gravito-magnetic charge ($N$) and/or the dual (magnetic) mass ($\tM = N$), together with a new
secondary hair ($J_N = MN$), namely, a Kerr-like conserved `angular momentum', enter into the standard
forms of the first law and Bekenstein-Smarr mass formula. Distinguished from other recent attempts, our
consistent thermodynamic differential and integral mass formulas are both obtainable from a meaningful
Christodoulou-Ruffini-type squared mass formula of almost all of the four-dimensional NUT-charged
spacetimes. As an excellent consequence, the famous Bekenstein-Hawking one-quarter area-entropy relation
can be naturally restored not only in the Lorentzian sector and but also in the Euclidian counterpart
of the generic Taub-NUT-type spacetimes without imposing any constraint condition.

However, only purely electric-charged cases in the four-dimensional Einstein-Maxwell gravity theory
with a NUT charge have been addressed there. In this paper, we shall follow the simple, systematic
way proposed in that article to further investigate the dyonic NUT-charged case. It is shown that
the standard thermodynamic relations continue to hold true provided that no new secondary charge is
added, however, the so-obtained electrostatic and magneto-static potentials are not coincident with
those computed via the standard method. To rectify this inconsistence, a simple strategy is provided
by further introducing two additional secondary hairs: $Q_N = QN$ and $P_N = PN$, together with their
thermodynamical conjugate potentials, so that the first law and Bekenstein-Smarr mass formula are still
satisfied, where $Q$ and $P$ being the electric and magnetic charges, respectively.
\end{abstract}

\maketitle
%\end{CJK*}

%%%%%%%%%%%%%%%%%%%%%%
\section{Introduction}
%%%%%%%%%%%%%%%%%%%%%%

In recent years, thermodynamics of the four-dimensional Lorentzian Taub-NUT spacetimes in the
Einstein-Maxwell gravity theory have attracted a lot of attention \cite{PRD100-064055,PLB798-134972,
JHEP0719119,JHEP0520084,CQG36-194001,PLB802-135270,1908.04238,PRD100-104016,JHEP0821152,2109.07715,
2112.00780,PRD101-124011}. In particular, in our previous work \cite{PRD100-101501}, we have advocated
a new idea that ``the NUT charge is a thermodynamical multi-hair" and put forward a simple, systematic
way to study the consistent thermodynamics of almost all of the four-dimensional NUT-charged spacetimes.
The consistent first law and Bekenstein-Smarr mass formula of these NUT-charged spacetimes are deduced
by first deriving a new meaningful Christodoulou-Ruffini-type squared-mass formula satisfied by the
four-dimensional NUT-charged spacetimes with a new secondary hair: $J_N = MN$. By contrast, it should
be mentioned that there is no analogous expression of the Christodoulou-Ruffini-type squared-mass
formula \cite{PRL25-1596,PRD4-3552} in all of the previous works \cite{PRD100-064055,PLB798-134972,
JHEP0719119,JHEP0520084,CQG36-194001,PLB802-135270,1908.04238,PRD100-104016,JHEP0821152,2109.07715,
2112.00780,PRD101-124011}. As a fact that has already been demonstrated in Ref. \cite{PLB634-531},
our new secondary hair $J_N = MN \equiv M_5$ exactly corresponds to the mass of the five-dimensional
gravitational magnetic monopole, so at least from the five-dimensional point of view, it is very natural
to consider it as a global conserved charge and then it is reasonable to include it to the first law
and Bekenstein-Smarr mass formula. There are many reasons to support such an idea. For instance,
it helps to explain the gyromagnetic ratio of Kerr-NUT-type spacetime \cite{PRD77-044038}, and the
quantization condition for a gravitational monopole \cite{CQG3-65,PPS92-1,GRG5-603}. What's more,
it is proved in Ref. \cite{PLB807-135521} that only considering the secondary hair $J_N = MN$ as
a independent charge, can the area (or entropy) products of the NUT-charged spacetimes be subject
to the universal rules \cite{PRL106-121301}, and the mass be expressed as a sum of the surface energy,
the rotational energy and the electromagnetic energy \cite{GRG53-69}.

According to the scheme advocated in our previous paper \cite{PRD100-101501}, the traditional elegant
Bekenstein-Hawking one-quarter area-entropy relation can be naturally restored in the Lorentzian and
Euclidian sectors of the generic NUT-charged spacetimes (and all of their extensions) in the
four-dimensional Einstein-Maxwell gravity theory without imposing any constraint condition. Due to
the fact that the NUT charge not only acts as a dual (magnetic) mass, but also simultaneously has
the rotation-like and electromagnetic charge-like characters, we arrive at a new recognition that
it must be a thermodynamical multi-hair. This viewpoint is in sharp contrast with all previous
knowledge that it has merely one physical feature, or that it is purely a single solution-parameter,
what is more, the physical meaning of the NUT parameter as a poly-facet can be completely uncovered
in the thermodynamical sense.

The four-dimensional NUT-charged spacetimes studied in our previous work \cite{PRD100-101501} are
either static charged (including a nonzero negative cosmological constant) or rotating charged (with
a vanishing cosmological constant) in the Einstein-Maxwell theory with a purely electric charge.
Note that the purely magnetic-charged case can be identically treated via the electric-magnetic
duality relation. However, that paper didn't consider the case of the four-dimensional dyonic
NUT-charged spacetimes, nor it dealt with the higher-dimensional case \cite{CQG19-2051,CQG23-2849,
CQG21-2937,PLB634-448,PLB632-537,CQG23-5323,NPB762-38} and those four-dimensional NUT-charged spacetimes
beyond the Einstein-Maxwell theory (such as: Kaluza-Klein (K-K) theory \cite{PRD103-064045,PRD77-124022,
CQG28-032001}, Einstein-Maxwell-Dilaton-Axion (EMDA) theory \cite{2111.06111} and more general (gauged)
STU supergravity theory \cite{NPB717-246,CQG31-022001,PRD90-025029}), all of which need to be studied
promptly. In the present paper, we shall focus on the thermodynamics of the four-dimensional Lorentzian
dyonic NUT-charged spacetimes in the Einstein-Maxwell gravity theory also.

The remaining part of this paper is organized as follows. In Sec. \ref{II}, we begin with a brief
introduction of some aspects of the four-dimensional Lorentzian dyonic Reissner-Nordstr\"{o}m-NUT
(RN-NUT) solution and then construct a new Christodoulou-Ruffini-like squared-mass formula, from
which both the differential and integral mass formulas can be derived via a simple mathematical
manipulation by only including the secondary hair $J_N = MN$, as did before in Ref. \cite{PRD100-101501}.
However, there exists a contradiction between the obtained electrostatic and magneto-static potentials
with those computed by the standard method. We demonstrated that this inconsistence can be simply
remedied by further introducing two new additional secondary hairs: $Q_N = QN$ and $P_N = PN$,
together with their thermodynamical conjugate potentials, where $Q$ and $P$ being the electric
and magnetic charges, respectively, so that the standard thermodynamic relations can continue to
hold true. In Sec. \ref{III}, we turn to discuss the case of the dyonic RN-NUT-AdS$_4$ spacetime.
We show that the dual (magnetic) mass must be further added to reproduce the familiar thermodynamical
volume delivered in other literatures. Then, in Sec. \ref{IV}, we extend the above work to the case
of the four-dimensional dyonic Kerr-Newman-NUT (KN-NUT) spacetime. In Sec. \ref{V}, we discuss the
impact of the secondary hair $J_N$ on the mass formulas and present the reduced mass formulas. Finally,
we present our conclusions in Sec. \ref{VI}.

%%%%%%%%%%%%%%%%%%%%%%%%%%%%%%%%%%%%%%%%%%%%%%%%%%%%%%%%%%%%%%%%%%%%%%%%%%%%%%%%%%%%%%%%%%%
\section{Consistent mass formulas of the four-dimensional dyonic RN-NUT spacetime}\label{II}
%%%%%%%%%%%%%%%%%%%%%%%%%%%%%%%%%%%%%%%%%%%%%%%%%%%%%%%%%%%%%%%%%%%%%%%%%%%%%%%%%%%%%%%%%%%

Let us start by summarizing some essential facts of the Lorentzian four-dimensional RN-NUT metric with
both electric and magnetic charges in the Lorentz sector \cite{JMP4-915,AP98-98}. We adopt the following
exquisite form of the line element in which the Misner strings \cite{JMP4-924} are symmetrically
distributed along the polar axis:
\bea
ds^2 &=& -\frac{f(r)}{r^2 +N^2}(dt +2N\cos\theta\, d\phi)^2 +\frac{r^2 +N^2}{f(r)}dr^2 \nn \\
&& +(r^2 +N^2)(d\theta^2 +\sin^2\theta\, d\phi^2) \, , \label{nut}
\eea
where $f(r) = r^2 -2Mr -N^2 +Q^2 +P^2$, in which $M$, $N$, $Q$ and $P$ are the mass, the NUT charge,
the electric and magnetic charges of the spacetime, respectively. In addition, the electro-magnetic
gauge potential one-form and its dual one-form are:
\bea
\mathbf{A} &=& \frac{Qr -PN}{r^2 +N^2}(dt +2N\cos\theta\, d\phi) +P\cos\theta\, d\phi \, , \label{A} \\
\widetilde{\mathbf{A}} &=& \frac{Pr +QN}{r^2 +N^2}(dt +2N\cos\theta\, d\phi)
 -Q\cos\theta\, d\phi \label{tA} \, ,
\eea
in which a gauge choice is made to let the temporal components of both potentials (\ref{A}, \ref{tA})
be zero at infinity, so that the corresponding electrostatic and magneto-static potentials vanish at
infinity. Alternatively, another often-used expressions for them are given by \cite{JHEP0719119,
PR133-B845}:
\bea
\mathbb{A} &=& \frac{2QNr +P(r^2-N^2)}{2N(r^2 +N^2)}(dt +2N\cos\theta\, d\phi) \, , \nn \\
\widetilde{\mathbb{A}} &=& \frac{2PNr -Q(r^2-N^2)}{2N(r^2 +N^2)}(dt +2N\cos\theta\, d\phi) \, , \nn
\eea
whose temporal components differ ours by two constants: $P/(2N), -Q/(2N)$, respectively.

Traditionally, the spacetime (\ref{nut}) is termed as being asymptotically local flat. It has a lot
of odd physical properties that are mainly due to the presence of the wire/line singularities at the
polar axis ($\theta = 0, \pi$), which are often dubbed the Misner strings, an analogue of the Dirac
string in electrodynamics. Misner \cite{JMP4-924} proposed to remove this kind of wire/line singularities
(so as to ensure the regularity of the metric) by imposing a time periodical identification condition:
$\beta = 8\pi n$. Then, the inevitable appearance of closed timelike curves subsequently led him
\cite{CWM} to claim that the NUT parameter was nonphysical and the Taub-NUT spacetime was ``a counter
example to almost anything" in General Relativity. However in recent years, Cl\'ement \textit{et al.}
\cite{PLB750-591,PRD93-024048,GRG50-60} demonstrated that actually it is not necessary to remove the
Misner string by imposing a periodicity condition of the time coordinate. They illustrated that the
Misner string singularities are far less problematic than previously thought, and argued that the
Lorentzian Taub-NUT solutions without the Misner time periodicity condition are geodesically complete,
and causality is not violated at all for geodesic observers, despite the existence of regions with
closed timelike curves. An immediate consequence of their researches is that the Lorentzian Taub-NUT
spacetimes with the Misner strings may be physical in nature. This, in turn, invokes a lot of recent
enthusiasm to explore other properties of these NUT-charged spacetimes.

In the following, we will derive various mass formulas and discuss the consistent thermodynamics
of the four-dimensional Lorentzian dyonic RN-NUT spacetime. As did in Refs. \cite{PRD100-064055,
PLB798-134972,1908.04238,PLB802-135270,PCPS66-145,PCPS70-89,CQG22-3555,CQG23-4473,PLB750-591,
PRD100-101501}, we will not impose the time periodicity condition, in the meanwhile, we shall
also keep the Misner strings symmetrically present at the polar axes and only concern about the
conical singularities satisfying $f(r) = 0$, namely, the outer and inner horizons located at
$r_h = r_{\pm} = M \pm\sqrt{M^2 +N^2 -Q^2 -P^2}$. Below, we will focus on the (exterior) event
horizon, and the discussions are also valid for the (interior) Cauchy horizon.

To begin with, let's recall some known quantities that can be evaluated via the standard method.
First, the area and the surface gravity at the horizon are easily computed as
\be
A_h = 4\pi(r_h^2 +N^2) = 4\pi\cA_h \, , \quad \kappa = \frac{f^{\prime}(r_h)}{2\cA_h}
 = \frac{r_h -M}{r_h^2 +N^2} \, , \quad \label{Ak}
\ee
with a `reduced horizon area' $\cA_h$ being introduced \cite{PRD100-101501,PLB608-251} just for
the later shortness:
\be
\cA_h = r_h^2 +N^2 = 2Mr_h +2N^2 -Q^2 -P^2 \, . \label{area}
\ee

The electrostatic and magneto-static potentials are gauge independent, by virtue of the above
specific gauge choice, they are simply given by
\be\ba
&\Phi = \Phi_h = (\mathbf{A}_{\mu}\xi^{\mu})|_{r=r_h} = \frac{Qr_h -PN}{r_h^2 +N^2} \, , \\
&\Psi = \Psi_h = (\widetilde{\mathbf{A}}_{\mu}\xi^{\mu})|_{r=r_h}
 = \frac{Pr_h +QN}{r_h^2 +N^2} \, , \label{emP}
\ea\ee
where $\xi = \p_t$ is a timelike Killing vector normal to the horizon.

As far as the calculation of the global conserved charges ($M, N, Q, P$) is concerned, the mass
$M$ can be computed via the Komar integral related to the timelike Killing vector $\p_t$, while
the electric and magnetic charges ($Q, P$) can be integrated by using the Gauss' law associated
with the field strengths ($F = d\mathbf{A}, \widetilde{F} = d\widetilde{\mathbf{A}}$), respectively.
The NUT charge $N$, however, has several different meanings, and so can be evaluated via different
methods. If it appears as the dual or magnetic-type mass \cite{JMP22-2612,JMP23-2168,DN1966,GRG5-603},
then it can be determined via the dual Komar integral as $\widetilde{M} = N$. On the other hand,
if it acts as the gravito-magnetic charge, it can be calculated via the definition given in Ref.
\cite{PRD59-024009}. One cannot distinguish the dual or magnetic-type mass from the gravito-magnetic
charge in the present case; however, we shall see below that they are significantly different from
each other once a nonzero cosmological constant is included. In addition, the conserved charges
($M, N, Q, P$) as the primary hairs apparently appear in the leading order of the asymptotic
expansions of the following components of the metric and the Abelian potentials at infinity:
\be\begin{split}
& g_{tt} \simeq -1 +\frac{2M}{r} +\mathcal{O}(r^{-2}) \, , \\
& g_{t\phi} \simeq \Big(-2N +\frac{4MN}{r}\Big)\cos\theta +\mathcal{O}(r^{-2}) \, , \\
& \mathbf{A}_t \simeq \frac{Q}{r} +\mathcal{O}(r^{-2}) \, , \qquad
\widetilde{\mathbf{A}}_t \simeq \frac{P}{r} +\mathcal{O}(r^{-2}) \, , \\
& \mathbf{A}_{\phi} \simeq \Big(P +\frac{2QN}{r}\Big)\cos\theta +\mathcal{O}(r^{-2}) \, , \\
& \widetilde{\mathbf{A}}_{\phi} \simeq \Big(-Q +\frac{2PN}{r}\Big)\cos\theta +\mathcal{O}(r^{-2}) \, .
\end{split}\ee
Note that our previously included secondary hair: $J_N = MN$ appears as the next leading order of the
asymptotic expansion of the metric component $g_{t\phi}$, and we can find that there are also two same
next leading order quantities $QN$ and $PN$ in the asymptotic expansions of two Abelian potentials'
components $\mathbf{A}_{\phi}$ and $\widetilde{\mathbf{A}}_{\phi}$, indicating that as two secondary
hairs, they might play an key role in the mass formula also.

\bigskip
%%%%%%%%%%%%%%%%%%%%%%%%%%%%%%%%%%%%%%%%%%%%%%%%%%%%%%%%%%%%%%%%%%%%%%%%%%%%%%%
\subsection{Mass formulas with the secondary hair: $J_N = MN$ only}\label{II.A}
%%%%%%%%%%%%%%%%%%%%%%%%%%%%%%%%%%%%%%%%%%%%%%%%%%%%%%%%%%%%%%%%%%%%%%%%%%%%%%%

In order to derive the first law which is reasonable and consistent in both physical and mathematical
sense, we adopt the method used in Refs. \cite{PRD100-101501,PLB608-251} to deduce a meaningful
Christodoulou-Ruffini-type squared mass formula. First, we rewrite the expression (\ref{area})
of the reduced horizon area and get the following identity:
\be
\big(\cA_h -2N^2 +Q^2 +P^2\big)^2 = 4M^2r_h^2 = 4M^2\cA_h -4M^2N^2 \, . \label{Sqarea}
\ee
Next, supposed that we only need to introduce the secondary hair: $J_N = MN$, as did in our
previous work \cite{PRD100-101501}, then we can obtain a useful identity:
\be
M^2 = \frac{1}{4\cA_h}\big(\cA_h -2N^2 +P^2 +Q^2\big)^2 +\frac{J_N^2}{\cA_h} \, , \label{sqm}
\ee
which is a Christodoulou-Ruffini-like squared-mass formula for the four-dimensional dyonic RN-NUT
spacetime. We point out that this formula (\ref{sqm}) consistently reduces to the one obtained in
the case of the four-dimensional RN-NUT spacetime \cite{PRD100-101501} when the magnetic charge
$P$ is turned off.

Below, we will derive the differential and integral mass formulas for the dyonic RN-NUT spacetime,
supposing that the primary hairs are the mass $M$, the NUT charge $N$, the electric and magnetic
charges ($Q, P$) as well as the only one secondary hair: $J_N = MN$. Given that the secondary hair
$J_N$ can be viewed temporarily as a independent variable\footnote{However, one may think that it
actually is not independent. A careful discussion about its impact on the mass formulas is presented
in Sec. \ref{V}.} at this moment, then the above squared-mass formula (\ref{sqm}) can be viewed
formally as a fundamental functional relation: $M = M(\cA_h, N, J_N, Q, P)$. Differentiating it
(multiplied by $4\cA_h$) with respect to the thermodynamical variables ($\cA_h, N, J_N, Q, P$)
yields their conjugate quantities, as was done in Refs. \cite{PLB608-251,PRD103-044014,PRD101-024057,
PRD102-044007,PRD21-884,CQG17-399,CPL23-1096}. In doing so, we can arrive at the differential and
integral mass formulas, with the conjugate thermodynamic potentials given by the ordinary Maxwell
relations.

Let us now demonstrate the above conclusion in more detail. Differentiating the squared-mass formula
(\ref{sqm}) with respect to the reduced horizon area $\cA_h$ yields one half of the surface gravity:
\bea
\kappa &=& 2\frac{\p\,M}{\p\,\cA_h}\Big|_{(N,J_N,Q,P)} = \frac{\cA_h -2N^2 +Q^2 +P^2 -2M^2}{2M\cA_h} \nn \\
&=& \frac{r_h -M}{r_h^2 +N^2} \, ,
\eea
which is entirely identical to the one given in Eq. (\ref{Ak}). Similarly, by the differentiation of
the squared-mass formula (\ref{sqm}) with respect to the NUT charge $N$ and the secondary hair $J_N$,
one can obtain the conjugate gravito-magnetic potential $\psi_h$ and the conjugate ``quasi-angular
momentum" $\omega_h$ as:
\bea
\psi_h &=& \frac{\p\,M}{\p\,N}\Big|_{(\cA_h,J_N,Q,P)} = \frac{-N(\cA_h -2N^2 +Q^2 +P^2)}{M\cA_h} \nn \\
&=& \frac{-2Nr_h}{r_h^2+N^2} \, , \\
\omega_h &=& \frac{\p\,M}{\p\,J_N}\Big|_{(\cA_h,N,Q,P)} = \frac{J_N}{M\cA_h} = \frac{N}{r_h^2+N^2} \, .
\eea
The electrostatic and magneto-static potentials, which are conjugate to $Q$ and $P$, respectively,
can be computed as:
\bea
\hat{\Phi} &=& \frac{\p\,M}{\p\,Q}\Big|_{(\cA_h,N,J_N,P)}
 = \frac{Q(\cA_h -2N^2 +Q^2 +P^2)}{2M\cA_h} \nn \\
&=& \frac{Qr_h}{r_h^2+N^2} \, , \\
\hat{\Psi} &=& \frac{\p\,M}{\p\,P}\Big|_{(\cA_h,N,J_N,Q)}
 = \frac{P(\cA_h -2N^2 +Q^2 +P^2)}{2M\cA_h} \nn \\
&=& \frac{Pr_h}{r_h^2+N^2} \, ,
\eea
which coincide with their corresponding ones only in the purely electric- or purely magnetic-charged
case \cite{PRD100-101501}. In the present dyonic case, these two quantities are apparently different
from those given in Eq. (\ref{emP}). Nevertheless, we can verify that both the differential and integral
mass formulas are completely satisfied
\bea
dM &=& (\kappa/2)\, d\cA_h +\omega_h\, dJ_N +\psi_h\, dN
 +\hat{\Phi}dQ +\hat{\Psi}dP \, , \quad \label{dmf} \\
M &=& \kappa\cA_h +2\omega_h\, J_N +\psi_h\, N +\hat{\Phi}Q +\hat{\Psi}P \, , \label{imf}
\eea
with respect to all the above thermodynamical conjugate pairs.

It is worth mentioning that the above differential and integral mass formulas (\ref{dmf}-\ref{imf})
can not only naturally reduce to the purely electric- or purely magnetic-charged case when the magnetic
or electric charge vanishes ($P = 0$ or $Q = 0$), but also smoothly recover the dyonic RN black hole
case when the NUT charge vanishes ($N = 0$). Comparing our new mass formulas presented in Eqs.
(\ref{dmf}-\ref{imf}) with the standard ones, it is strongly suggested that one should make the
following familiar identifications:
\be
T = \frac{\kappa}{2\pi} = \frac{r_h -M}{2\pi(r_h^2 +N^2)} \, , \quad
S = \frac{A}{4} = \pi(r_h^2 +N^2) \, , \label{ts}
\ee
which restores the famous Bekenstein-Hawking one-quarter area-entropy relation of the dyonic RN-NUT
spacetime in a very comfortable way. It is worth noting that one should assign a geometric entropy
to the dyonic RN-NUT spacetime, which is just one-quarter of its horizon area. In the above `derivation',
we do not require in advance that the relation (\ref{ts}) must hold in order to obtain a reasonable
first law, but rather it is a very natural result from the above thermodynamic derivation.

It is remarkable that unlike Ref. \cite{JHEP0719119}, our differential and integral mass formulas
(\ref{dmf})-(\ref{imf}) attain their traditional forms which relate the global conserved charges
($M, Q, P, N, J_N$) measured at the infinity to those quantities ($T, S, \hat{\Phi}, \hat{\Psi},
\psi_h, \omega_h$) evaluated at the horizon. In this sense, it is quite reasonable to infer that
the entire set of four laws of the usual black hole thermodynamics is completely applicable to the
dyonic RN-NUT spacetime. It's time to formally call the dyonic NUT-charged spacetimes as real black
holes, at least from the thermodynamic point of view.

%%%%%%%%%%%%%%%%%%%%%%%%%%%%%%%%%%%%%%%%%%%%%%%%%%%%%%%%%%%%%%
\subsection{Two new secondary hairs $Q_N = QN$ and $P_N = PN$}
%%%%%%%%%%%%%%%%%%%%%%%%%%%%%%%%%%%%%%%%%%%%%%%%%%%%%%%%%%%%%%

In the last subsection, we have derived the differential and integral mass formulas of the
four-dimensional dyonic RN-NUT spacetime via differentiating the squared-mass formula (\ref{sqm}),
but with a fly in the ointment as mentioned earlier, namely, the derived expressions for the
conjugate electrostatic and magneto-static potentials are inconsistent with those previously
calculated by using the standard method. In order to get a consistent and reasonable first law
and Bekenstein-Smarr mass formula, this inconsistence must be removed now. Noting that the
expressions $\psi_h = -2Nr_h/(r_h^2+N^2)$ and $\omega_h = N/(r_h^2+N^2)$ in the mass formulas
(\ref{dmf}-\ref{imf}) do not explicitly contain the electric and magnetic charges ($Q, P$),
so we can leave them unchanged and replace only the electrostatic and magneto-static potentials
($\hat{\Phi}, \hat{\Psi}$) by the standard ones ($\Phi, \Psi$) given in Eq. (\ref{emP}). First, using
$\hat{\Phi} Q + \hat{\Psi} P = \Phi Q + \Psi P$, the integral mass formula (\ref{imf}) can be rewritten as
\be
M = 2T S +2\omega_h\, J_N +\psi_h\, N +\Phi\, Q +\Psi\,P \, . \label{BS}
\ee
Next, the first law (\ref{dmf}) can be rewritten as
\bea
dM &=& T dS +\omega_h\, dJ_N  +\psi_h\, dN +\Phi\, dQ +\Psi\, dP \nn \\
&& +\frac{N}{r_h^2 +N^2}(PdQ -QdP) \, , \nn \\
&=& T dS +\omega_h\, dJ_N  +\psi_h\, dN +\Phi\, dQ +\Psi\, dP \nn \\
&& +\frac{Pd(QN) -Qd(PN)}{r_h^2 +N^2} \, , \nn \\
&=& T dS +\omega_h\, dJ_N +\psi_h\, dN +\Phi\, dQ +\Psi\, dP \nn \\
&& +\Phi_N\, dQ_N +\Psi_N\, dP_N \, , \label{FL}
\eea
provided that one further introduces two new additional secondary hairs: $Q_N = QN$ and $P_N = PN$,
together with their thermodynamic conjugate potentials:
\be
\Phi_N = \frac{P}{r_h^2 +N^2} \, , \quad \Psi_N = \frac{-Q}{r_h^2 +N^2} \, .
\ee
Also, since $\Phi_N Q_N + \Psi_N P_N = 0$, so the Bekenstein-Smarr mass formula (\ref{BS}) can be
further rewritten as
\be
M = 2TS + 2\omega_h J_N + \psi_h N + \Phi\, Q +\Psi\, P +\Phi_N Q_N + \Psi_N P_N\, .
\ee
From the first law (\ref{FL}), it is easy to see that there are five cases
with no need to introduce the secondary hairs: $Q_N = QN$ and $P_N = PN$ as well as their conjugate
potentials ($\Phi_N, \Psi_N$): i) purely electric-charged case ($P = 0$); ii) purely magnetic-charged
case ($Q = 0$); iii) dyonic RN solution ($N = 0$); iv) self-dual vector potential case ($Q = P$); and
v) anti-self-dual vector potential case ($Q = -P$).

The above identities (\ref{BS}-\ref{FL}) are the expected standard forms of our consistent first law
and Bekenstein-Smarr mass formula for the dyonic RN-NUT spacetime, suggesting that the NUT charge
should be treated as a thermodynamic multi-hair. The advantage of introducing the above secondary
hairs is as follows: 1) it can smoothly recover the cases where the solution parameters take some
special values in our previous work \cite{PRD100-101501}; 2) it can retain some thermodynamic quantities
calculated by the standard method also; 3) all the expressions of the related thermodynamic quantities
are very concise and much more simple than those appeared in other literatures.

Finally, if the squared-mass (\ref{sqm}) is viewed as a binomial of the reduced horizon area:
\bea
\cA_h^2 && +2(P^2+Q^2-2N^2-2M^2)\cA_h \nn \\
&& +4M^2N^2 +(P^2+Q^2-2N^2)^2 = 0 \, ,
\eea
then the area product of the inner and outer horizons
\be
\cA_+\cA_- = 4J_N^2 +(P^2+Q^2-2N^2)^2 \,
\ee
can be quantized only when $J_N = MN$ is quantized in a manner like the quantization of the angular
momentum, and the charges ($Q, P, N$) take some discrete values also.

%%%%%%%%%%%%%%%%%%%%%%%%%%%%%%%%%%%%%%%%%%%%%%%%%%%%%%%%%%%%%%%%%%%%%
\section{Extension to the dyonic RN-NUT-AdS$_4$ spacetime}\label{III}
%%%%%%%%%%%%%%%%%%%%%%%%%%%%%%%%%%%%%%%%%%%%%%%%%%%%%%%%%%%%%%%%%%%%%

In this section, we would like to extend the above work to the Lorentzian dyonic RN-NUT-AdS$_4$ spacetime
with a nonzero cosmological constant. The metric, the Abelian gauge potential and its dual are still
given by Eqs. (\ref{nut}-\ref{tA}), but now $f(r) = r^2 -2Mr -N^2 +Q^2 +P^2 +g^2(r^4 +6N^2r^2 -3N^4)$,
in which $g = 1/l$ is the gauge coupling constant.

First, we will determine the conserved charges (primary hairs) of the dyonic RN-NUT-AdS$_4$ solution.
The electric and magnetic charges ($Q, P$) as well as the gravito-magnetic charge $N$ can be computed
just like the case without a cosmological constant. We will adopt the conformal completion method to
calculate its electric mass $M$ and dual (magnetic) mass \cite{PRD93-084022}, and show that the dual
(magnetic) mass $\widetilde{M}$ is different from the NUT charge $N$ now. The conformal boundary metric
of the dyonic RN-NUT-AdS$_4$ spacetime is given by:
\bea
ds_\infty^2 &=& \lim_{r\to\infty}\frac{ds^2}{r^2} = -g^2(dt +2N\cos\theta d\phi)^2 \nn \\
&&\qquad\qquad +d\theta^2 +\sin^2\theta d\phi^2 \, ,
\eea
with $g^{rr} = g^2r^4$ being used to define a normal vector $N^r = gr^2$, then the conserved charge
$\mathcal{Q}[\xi]$ associated with the Killing vector $\xi = \p_t$ is given by
\be
\mathcal{Q}[\xi] = \frac{1}{8\pi g^3}\int\, rN^\alpha\, N^\beta\,
 C^{\mu}_{~\alpha\nu\beta}\xi^{\nu}dS_\mu \, ,
\ee
where $C^{\mu}_{~\alpha\nu\beta}$ is the Weyl conformal tensor and
\be
dS_\mu = g\sin\theta\, d\theta \wedge\, d\phi
\ee
is the area element of the 2-spherical cross section of the conformal boundary. The conformal
(electric) mass $\mathcal{M}$ is easily evaluated as:
\be
\mathcal{M} = \mathcal{Q}[\xi^t] = M \, .
\ee

Similarly, in order to evaluate the dual conformal mass, we can define a dual conserved charge
$\widetilde{\mathcal{Q}}[\xi]$ via replacing the Weyl conformal tensor by its left-dual:
\be
\widetilde{C}_{\mu\nu\rho\sigma} = \frac{1}{2}\epsilon_{\mu\nu\alpha\beta}
 C^{\alpha\beta}_{~~~\rho\sigma} \, ,
\ee
where $\epsilon_{\mu\nu\alpha\beta}$ is the Levi-Civita antisymmetry tensor. Then, the dual
(magnetic) mass is computed as
\be
\widetilde{M} = \widetilde{\mathcal{Q}}[\xi^{t}] = N(1 +4g^2N^2) \, ,
\ee
which is unequal to the NUT charge $N$.

Next, we want to calculate some thermodynamic quantities associated with the Killing horizons
defined by $f(r_h) = 0$. The surface gravity at the horizon is given by
\be
\kappa = \frac{f^{\prime}(r_h)}{2\cA_h} = \frac{r_h -M +2g^2(r_h^2 +3N^2)r_h}{\cA_h} \, ,
\label{Ak1}
\ee
while the horizon area still reads: $A_h = 4\pi(r_h^2 +N^2) = 4\pi\cA_h$ in which the reduced
horizon area now is
\bea
\cA_h &=& 2Mr_h +2N^2 -Q^2 -P^2 -g^2(r_h^4 +6N^2r_h^2 -3N^4) \nn \\
&=& 2Mr_h +2N^2 -Q^2 -P^2 \nn \\
&& -g^2(\cA_h^2 +4N^2\cA_h -8N^4) \, . \label{rha}
\eea
The electrostatic and magneto-static potentials are
\be
\Phi = \frac{Qr_h -PN}{r_h^2 +N^2} \, , \quad \Psi = \frac{Pr_h +QN}{r_h^2 +N^2} \, , \label{PP1}
\ee
which own the same expressions as those given in Eq. (\ref{emP}), although the horizon location
$r_h$ now has a different expression.

\bigskip
%%%%%%%%%%%%%%%%%%%%%%%%%%%%%%%%%%%%%%%%%%%%%%%%%%%%%%%%%%%%%%%%%%%%%%%%%%%%%%%%
\subsection{Mass formulas with the secondary hair: $J_N = MN$ only}\label{III.A}
%%%%%%%%%%%%%%%%%%%%%%%%%%%%%%%%%%%%%%%%%%%%%%%%%%%%%%%%%%%%%%%%%%%%%%%%%%%%%%%%

Now, we assume that only the secondary hair $J_N = MN$ is needed as before, and deduce a squared-mass
formula. The reduced horizon area (\ref{rha}) can be collected as:
\be
2Mr_h = (1 +4g^2N^2)(\cA_h -2N^2) +Q^2 +P^2 +g^2\cA_h^2 \, ,
\ee
after squaring this identity and then adding $4M^2N^2$ to its left hand side and $4J_N$ to its right
hand side, we obtain
\bea
M^2 &=& \frac{1}{4\cA_h}\big[(1 +4g^2N^2)(\cA_h -2N^2) \nn \\
&& +Q^2 +P^2 +g^2\cA_h^2\big]^2 +\frac{J_N^2}{\cA_h} \, ,
\eea
which is nothing but the squared-mass formula
\bea
M^2 &=& \frac{1}{\cA_h}\bigg[\Big(1 +\frac{32\pi}{3}\cP N^2 \Big)(\cA_h -2N^2) \nn \\
&& +Q^2 +P^2 +\frac{8\pi}{3}\cP\cA_h^2 \bigg]^2 +\frac{J_N^2}{\cA_h} \, , \label{Sqm}
\eea
after introduce the generalized pressure $\cP = 3g^2/(8\pi)$ to replace the cosmological constant.
We also point out that the squared-mass formula (\ref{Sqm}) consistently reduces to the one obtained
in the four-dimensional RN-NUT-AdS spacetime case \cite{PRD100-101501} when the magnetic charge $P$
vanishes.

In the following, the differential and integral mass formulas for the dyonic RN-NUT-AdS$_4$ spacetime
are derived by assuming that the whole set of thermodynamic quantities is the mass $M$, the NUT charge
$N$, the electric and magnetic charges ($Q, P$), the generalized pressure $\cP$ and the only one
secondary hair: $J_N = MN$ which will be viewed as a independent variable also.\footnote{A detailed
discussion about the impact of $J_N = MN$ on the mass formulas is presented in Sec. \ref{V}} In this
way, the squared-mass formula (\ref{Sqm}) then can be viewed as a fundamental functional relation:
$M = M(\cA_h, N, J_N, Q, P, \cP)$ of its thermodynamical variables.

Applying a similar procedure as manipulated in the last section, that is, performing the partial
derivative of the above squared-mass formula (\ref{Sqm}) (multiplied by $4\cA_h$) with respect to
one of the thermodynamical quantities ($\cA_h, N, J_N, Q, P, \cP$) and simultaneously fixing the
remaining ones, respectively, leads to its corresponding conjugate quantities. First, differentiating
the squared-mass formula (\ref{Sqm}) with respect to the reduced horizon area $\cA_h$ yields one
half of the surface gravity:
\be
\kappa = 2\frac{\p\,M}{\p\,\cA_h}\Big|_{(N,J_N,Q,P,\cP)}
 = \frac{r_h -M +2g^2(r_h^2 +3N^2)r_h}{r_h^2 +N^2} \, ,
\ee
which coincides with the one given in Eq. (\ref{Ak1}). Next, the potential $\psi_h$ and the
``quasi-angular momentum" $\omega_h$, which are conjugate to $N$ and $J_N$, respectively, are
given by
\bea
\psi_h &=& \frac{\p\,M}{\p\,N}\Big|_{(\cA_h,J_N,Q,P,\cP)}
 = 2Nr_h\frac{-1 +2g^2(r_h^2 -3N^2)}{r_h^2 +N^2} \, , \quad \\
\omega_h &=& \frac{\p\,M}{\p\,J_N}\Big|_{(\cA_h,N,Q,P,\cP)} = \frac{N}{r_h^2 +N^2} \, .
\eea
By differentiating the squared-mass formula (\ref{Sqm}) with respect to the electric and magnetic
charges ($Q, P$), respectively, one can get the conjugate electrostatic and magneto-static potentials
as follows:
\bea
\hat{\Phi} &=& \frac{\p\,M}{\p\,Q}\Big|_{(\cA_h,N,J_N,P,\cP)} = \frac{Qr_h}{r_h^2+N^2} \, , \\
\hat{\Psi} &=& \frac{\p\,M}{\p\,P}\Big|_{(\cA_h,N,J_N,Q,\cP)} = \frac{Pr_h}{r_h^2+N^2} \, .
\eea
These two quantities are also different from those given in Eq. (\ref{PP1}). Finally, via the
differentiation of the squared-mass formula (\ref{Sqm}) with respect to the pressure $\cP$, one
can obtain a conjugate thermodynamical volume:
\be
\cV = \frac{\p\,M}{\p\,\cP}\Big|_{(\cA_h,N,J_N,Q,P)}
 = \frac{4\pi\, r_h(r_h^4 +6N^2r_h^2 -3N^4)}{3(r_h^2+N^2)} \, .
\ee

Using all of the above thermodynamical conjugate pairs, we can easily check that both differential
and integral mass formulas are completely obeyed:
\bea
dM &=& (\kappa/2)d\cA_h +\omega_h\, dJ_N +\psi_h\, dN +\cV d\cP \nn \\
&& +\hat{\Phi}dQ +\hat{\Psi}dP \, , \label{dmf1} \\
M &=& \kappa\cA_h +2\omega_h\, J_N +\psi_h \, N  -2\cV\cP
 +\hat{\Phi}Q +\hat{\Psi}P \, . \quad \label{imf1}
\eea

It is natural to recognize
\be
S = \frac{A_h}{4} = \pi\cA_h \, , \quad T = \frac{\kappa}{2\pi} = \frac{f^{\prime}(r_h)}{4\pi\cA_h}\, ,
\ee
so that the solution behaves like a genuine black hole without violating the beautiful one-quarter
area/entropy law. In sharp contrast with Refs. \cite{JHEP0719119}, here we do not require in advance
that the first law should be obeyed so as to obtain the consistent thermodynamical relations, rather
it is just a very natural by-product of the pure algebraic deduction.

%%%%%%%%%%%%%%%%%%%%%%%%%%%%%%%%%%%%%
\subsection{Consistent mass formulas}
%%%%%%%%%%%%%%%%%%%%%%%%%%%%%%%%%%%%%

One may notice that there are two shortcomings of our work done in the last section. The first one
is that the obtained electrostatic and magneto-static potentials do not coincide with those computed
via the standard method; and the second one is that our derived conjugate thermodynamical volume $\cV$
is not equal to the familiar one: $\widetilde{\cV} = 4\pi\, r_h(r_h^2 +3N^2)/3$, which appeared in
other literatures \cite{PRD100-064055,PLB798-134972,JHEP0719119,CQG36-194001,CQG31-235003}. Below,
we will one by one resolve these two inconsistences.

To settle down the first contradiction, likewise the case without a cosmological constant, we just
need to further introduce two new additional secondary hairs: $Q_N = QN$ and $P_N = PN$, together
with their thermodynamic conjugate electrostatic and magneto-static potentials:
\be
\Phi_N = \frac{P}{r_h^2 +N^2} \, , \quad \Psi_N = \frac{-Q}{r_h^2 +N^2} \, ,
\ee
to get the standard forms of the Bekenstein-Smarr formula and the first law as follows:
\bea
M &=& 2T S +2\omega_h\, J_N +\psi_h\, N +\Phi\, Q +\Psi\, P -2\cV\cP \, , \label{BS1} \\
dM &=& T dS +\omega_h\, dJ_N +\psi_h\, dN  +\Phi\, dQ +\Psi\, dP \nn \\
 && +\cV\, d\cP +\frac{Pd(QN) -Qd(PN)}{r_h^2 +N^2} \, , \nn \\
 &=& T dS +\omega_h\, dJ_N +\psi_h\, dN +\Phi\, dQ +\Psi\, dP \nn \\
&& +\cV\, d\cP +\Phi_N\, dQ_N +\Psi_N\, dP_N \, , \label{FL1}
\eea
which are our consistent and reasonable thermodynamical first law and Bekenstein-Smarr mass formula
for the dyonic RN-NUT-AdS$_4$ spacetime. The first law (\ref{FL1}) indicates that there are three
classes of special cases without introducing the secondary hairs: $Q_N = QN$ and $P_N = PN$ as well
as their conjugate potentials ($\Phi_N, \Psi_N$): i) purely electric-charged ($P = 0$) or purely
magnetic-charged ($Q = 0$) case; ii) NUT-less dyonic solution ($N = 0$); iii) self-dual or anti-self-dual
vector potential case ($|Q| = |P|$).

On the basis of this modification, now we are ready to remove the second conflict via replacing the
derived conjugate thermodynamical volume $\cV$ by $\widetilde{\cV} = 4\pi\, r_h(r_h^2+3N^2)/3$, and
further introducing the dual (magnetic) mass $\tM = N(1+4g^2N^2)$ into the above differential and
integral mass formulas (\ref{BS1}-\ref{FL1}). Now we get the following consistent mass formulas:
\bea
dM &=& TdS +\omega_h\, dJ_N +\tph\, dN +\zeta\, d\tM +\Phi\, dQ +\Psi dP \nn \\
&& +\Phi_N\, dQ_N +\Psi_N\, dP_N +\widetilde{\cV}\, d\cP \, , \nn \\
M &=& 2TS +2\omega_h\, J_N +\tph\, N +\zeta\tM +\Phi\, Q +\Psi dP -2\widetilde{\cV}\, \cP \, , \nn
\eea
in which two new conjugate potentials are given by:
\bea
\tph = -\frac{2Nr_h}{\cA_h} -(1 -4g^2N^2)\zeta\, , \quad
\zeta = \frac{r_h(r_h^2-3N^2)}{4N\cA_h} \, . \nn
\eea
It is of little possibility to reproduce the thermodynamical volume $\widetilde{\cV}$ without the
inclusion of the dual mass $\tM$.

It should be pointed out that unlike the formalism advocated in other papers \cite{JHEP0719119,
PRD100-104016} where there are electric-type, magnetic-type, mixed-type and even many other versions
of the `consistent' first law in which the thermodynamic mass also remains unmodified, here our
consistent mass formulas are unique. By contrast, Awad et al. \cite{PRD101-124011} proposed to modify
the thermodynamic mass which includes the contribution from the Misner string so that the first law
retains its usual form without introducing new thermodynamical conjugate pairs, although they used
a four-dimensional planar NUT-charged spacetime as a special example. According to this fashion,
it is shown in Appendix \ref{appA} that there are infinitely many `consistent' mass formulas for
the dyonic RN-NUT-AdS$_4$ spacetime.

%%%%%%%%%%%%%%%%%%%%%%%%%%%%%%%%%%%%%%
\section{Consistent mass formulas of
the dyonic KN-NUT spacetime}\label{IV}
%%%%%%%%%%%%%%%%%%%%%%%%%%%%%%%%%%%%%%

Finally, we will show that the general rotating Lorentzian dyonic NUT-charged case without a
cosmological constant can be treated completely in the same pattern as did in the last two
sections. The line element of the dyonic KN-NUT spacetime with the Misner strings symmetrically
distributed along the rotation axis, the electromagnetic one-form and its dual one-form are:
\bea
ds^2 &=& -\frac{\Delta(r)}{\Sigma} X^2 +\frac{\Sigma}{\Delta(r)}dr^2
 +\Sigma d\theta^2 +\frac{\sta2}{\Sigma}Y^2 \, , \\
\mathbf{A} &=& \frac{Qr -PN}{\Sigma} X -\frac{P\cos\theta}{\Sigma} Y \nn \\
&=& \frac{Qr -P(N +a\cos\theta)}{\Sigma} X +P\cos\theta\, d\phi \, , \\
\widetilde{\mathbf{A}} &=& \frac{Pr +QN}{\Sigma} X +\frac{Q\cos\theta}{\Sigma} Y \nn \\
&=& \frac{Pr +Q(N +a\cos\theta)}{\Sigma} X -Q\cos\theta\, d\phi \, ,
\eea
where $\Sigma =r^2 +(N +a\cos\theta)^2$, and
\bea
\Delta(r) &=& r^2 +a^2 -2Mr -N^2 +Q^2 +P^2 \, , \nn \\
X &=& dt +(2N\cos\theta -a\sin^2\theta)d\phi \, , \nn \\
Y &=& a\, dt -(r^2 +a^2 +N^2)d\phi \, . \nn
\eea

The global conserved charges for this spacetime are the Komar mass $M$, the angular momentum $J = Ma$,
the electric and magnetic charges ($Q, P$), and the gravito-magnetic charge or dual (magnetic) mass
(both of which are identical to the NUT charge $N$). These conserved charges display obviously in the
leading order of the following asymptotic expansions of the metric components and the Abelian potentials
at infinity:
\be\begin{split}
& g_{tt} \simeq -1 +\frac{2M}{r} +\mathcal{O}(r^{-2}) \, , \\
& \mathbf{A}_t \simeq \frac{Q}{r} +\mathcal{O}(r^{-2}) \, , \qquad
\widetilde{\mathbf{A}}_t \simeq \frac{P}{r} +\mathcal{O}(r^{-2}) \, , \\
& g_{t\phi} \simeq \Big(-2N +\frac{4MN}{r}\Big)\cos\theta
 -\frac{2Ma}{r}\sin^2\theta +\mathcal{O}(r^{-2}) \, , \\
& \mathbf{A}_{\phi} \simeq \Big(P +\frac{2QN}{r}\Big)\cos\theta
 -\frac{Qa}{r}\sin^2\theta  +\mathcal{O}(r^{-2}) \, , \\
& \widetilde{\mathbf{A}}_{\phi} \simeq \Big(-Q +\frac{2PN}{r}\Big)\cos\theta
 -\frac{Pa}{r}\sin^2\theta +\mathcal{O}(r^{-2}) \, .
\end{split}\ee
Besides these primary hairs, there are also the secondary hairs: ($MN, QN, PN$), electric-dipole and
magnetic-dipole moments ($Qa, Pa$) appearing in the next leading order of the above asymptotic expansions.

The event and Cauchy horizons are determined by $\Delta(r_h) = 0$, which gives $r_h =r_{\pm} = M \pm
\sqrt{M^2 +N^2 -Q^2 -P^2 -a^2}$. The event horizon area is: $A_h = 4\pi\cA_h$ with the reduced horizon
area now being: $\cA_h = r_h^2 +a^2 +N^2 = 2Mr_h +2N^2 -Q^2 -P^2$.

At the horizon, the surface gravity and the angular velocity can be evaluated via the standard method as
\be
\kappa = \frac{\Delta^{\prime}(r_h)}{2\cA_h} = \frac{r_h -M}{\cA_h} \, , \quad
\Omega = \frac{-g_{t\phi}}{g_{\phi\phi}}\bigg|_{r = r_h} = \frac{a}{\cA_h}  \, . \label{KO}
\ee
The electrostatic and magneto-static potentials simply identify with those at the horizons and read
\be\ba
&\Phi = \Phi_h = (\mathbf{A}_{\mu}\xi^{\mu})|_{r=r_h}
 = \frac{Qr_h -PN}{\cA_h} \, , \\
&\Psi = \Psi_h = (\widetilde{\mathbf{A}}_{\mu}\xi^{\mu})|_{r=r_h}
 = \frac{Pr_h +QN}{\cA_h} \, , \label{PP2}
\ea\ee
where $\xi = \p_t +\Omega \p_{\phi}$ is the co-rotating Killing vector normal to the horizon.

\bigskip
%%%%%%%%%%%%%%%%%%%%%%%%%%%%%%%%%%%%%%%%%%%%%%%%%%%%%%%%%%%%%%%%%%%%%%%%%%%%%%%
\subsection{Mass formulas with the secondary hair: $J_N = MN$ only}\label{IV.A}
%%%%%%%%%%%%%%%%%%%%%%%%%%%%%%%%%%%%%%%%%%%%%%%%%%%%%%%%%%%%%%%%%%%%%%%%%%%%%%%

Adopting the same procedure as did in the last two sections and supposing that only one secondary
hair: $J_N = MN$ is needed to be included as before, we square the following identity: $2Mr_h =
\cA_h -2N^2 +Q^2 +P^2$, then after adding $4M^2(a^2 +N^2)$ to its left hand side and $4J^2 +4J_N^2$
to its right hand side, followed by dividing both sides with $4M^2$, we can obtain a squared-mass
formula:
\be
M^2 = \frac{1}{4\cA_h}(\cA_h -2N^2 +Q^2 +P^2)^2 +\frac{J_N^2 +J^2}{\cA_h} \, . \label{SQM}
\ee
which consistently reduces to the one obtained in the four-dimensional KN-NUT spacetime case
\cite{PRD100-101501} when the magnetic charge $P$ is turned off.

Incidentally, if the squared-mass (\ref{SQM}) is rewritten as a binomial of the reduced horizon area:
\bea
\cA_h^2 && +2(P^2+Q^2-2N^2-2M^2)\cA_h \nn \\
&& +4J^2 +4M^2N^2 +(P^2+Q^2-2N^2)^2 = 0 \, ,
\eea
then the area product of the inner and outer horizons:
\be
\cA_+\cA_- = 4J^2 +4J_N^2 +(P^2+Q^2-2N^2)^2 \,
\ee
can be quantized only when $J_N = MN$ is quantized in a manner just as the angular momentum $J = m\hbar$
is quantized, and the charges ($Q, P, N$) take some discrete values in the meanwhile.

Supposed temporarily that the secondary hair: $J_N = MN$ is a independent conserved charge, namely,
it can be treated as a independent thermodynamical variable,\footnote{We will discuss the impact
of $J_N = MN$ in Sec. \ref{V}.} then Eq. (\ref{SQM}) formally represents a fundamental functional
relation: $M = M(\cA_h, J, J_N, N, Q, P)$ with the whole set of the extensive variables being the
NUT charge $N$, the electric and magnetic charges ($Q, P$), the angular momentum $J$, the secondary
hair $J_N$, and $\cA_h$ as the intense quantity of the dyonic KN-NUT spacetime. Then, differentiating
the above squared-mass formula (\ref{SQM}) with respect to one variable of the whole set of the
thermodynamical quantities ($\cA_h, J, J_N, N, Q, P$) and simultaneously fixing the remaining ones,
respectively, gives rise to its corresponding conjugate quantities. Subsequently, one can derive
the differential and integral mass formulas with the conjugate thermodynamical potentials reproduced
by the ordinary Maxwell relations.

The conjugate quantity of the reduced horizon area $\cA_h$ is one half of the surface gravity:
\be
\kappa = 2\frac{\p\,M}{\p\,\cA_h}\Big|_{(J,J_N,N,Q,P)} = \frac{r_h -M}{\cA_h} \, .
\ee
The angular velocity, which is conjugate to $J$, is given by
\be
\Omega = \frac{\p\,M}{\p\,J}\Big|_{(\cA_h,J_N,N,Q,P)} = \frac{a}{\cA_h} \, . \label{Omega}
\ee
These two conjugate quantities are entirely identical to those given in Eq. (\ref{KO}). Differentiating
the squared-mass formula (\ref{SQM}) with respect to the NUT charge $N$ and the secondary hair $J_N$,
one can get the conjugate gravito-magnetic potential:
\be
\psi_h = \frac{\p\,M}{\p\,N}\Big|_{(\cA_h,J,J_N,Q,P)} = -\frac{2Nr_h}{\cA_h} \, ,
\ee
and a conjugate ``quasi-angular momentum":
\be
\omega_h = \frac{\p\,M}{\p\,J_N}\Big|_{(\cA_h,J,N,Q,P)} = \frac{N}{\cA_h} \, .
\ee
Differentiating the squared-mass formula (\ref{SQM}) with respect to the electric and magnetic charges
($Q, P$), respectively, yields the conjugate electrostatic and magneto-static potentials:
\bea
\hat{\Phi} &=& \frac{\p\,M}{\p\,Q}\Big|_{(\cA_h,J,J_N,N,P)} = \frac{Qr_h}{\cA_h} \, , \\
\hat{\Psi} &=& \frac{\p\,M}{\p\,P}\Big|_{(\cA_h,J,J_N,N,Q)} = \frac{Pr_h}{\cA_h} \, ,
\eea
which are different from those given in Eq. (\ref{PP2}).

One can also easily demonstrate both the differential and integral mass formulas are completely
fulfilled
\bea
dM &=& (\kappa/2)d\cA_h +\Omega\, dJ +\omega_h\, dJ_N +\psi_h\, dN \nn \\
&& +\hat{\Phi}dQ +\hat{\Psi}dP \, , \quad \label{dmf2} \\
M &=& \kappa\cA_h +2\Omega\, J +2\omega_h\, J_N +\psi_h\, N
 +\hat{\Phi}Q +\hat{\Psi}P \, , \quad \label{imf2}
\eea
after using all the above thermodynamical conjugate pairs.

The consistency of the above mass formulas (\ref{dmf2}-\ref{imf2}) suggests that one should restore
the well-known Bekenstein-Hawking area/entropy relation and Hawking temperature
\be
S = \frac{A_h}{4} = \pi\cA_h \, , \quad T = \frac{\kappa}{2\pi} = \frac{r_h -M}{2\pi\cA_h} \, ,
\ee
which means that the whole class of the four-dimensional dyonic NUT-charged spacetimes should be viewed
as generic black holes.

%%%%%%%%%%%%%%%%%%%%%%%%%%%%%%%%%%%%%%%%%%%%%%%%%%%%%%%%%%%%%%
\subsection{Two new secondary hairs $Q_N = QN$ and $P_N = PN$}
%%%%%%%%%%%%%%%%%%%%%%%%%%%%%%%%%%%%%%%%%%%%%%%%%%%%%%%%%%%%%%

In this subsection, we will show that a consistent and reasonable first law and Bekenstein-Smarr
mass formula of the dyonic KN-NUT spacetime can be obtained still via introducing two new additional
secondary hairs: $Q_N = QN$ and $P_N = PN$, together with their thermodynamic conjugate potentials:
\be
\Phi_N = \frac{P}{\cA_h} \, , \quad \Psi_N = \frac{-Q}{\cA_h} \, .
\ee

By the replacement of the electrostatic and magneto-static potentials, it is not difficult to see
that the integral mass formula becomes
\be
M = 2T S +2\Omega\, J +2\omega_h\, J_N +\psi_h\, N +\Phi\, Q +\Psi\, P \, , \label{BS3}
\ee
while the differential mass formula is rewritten as follows:
\bea
dM &=& T dS +\Omega\, dJ +\omega_h\, dJ_N +\psi_h\, dN +\Phi\, dQ \nn \\
&& +\Psi\, dP +\frac{Pd(QN) -Qd(PN)}{\cA_h} \, , \nn \\
&=& T dS +\Omega\, dJ +\omega_h\, dJ_N +\psi_h\, dN +\Phi\, dQ \nn \\
&& +\Psi\, dP +\Phi_N\, dQ_N +\Psi_N\, dP_N \, . \label{FL3}
\eea
The first law (\ref{FL3}) implies that there are three kind of special cases with no need of introducing
the secondary hairs: $Q_N = QN$ and $P_N = PN$ as well as their conjugate potentials ($\Phi_N, \Psi_N$):
i) purely electric-charged ($P = 0$) or purely magnetic-charged ($Q = 0$) case; ii) dyonic KN solution
($N = 0$); iii) self-dual or anti-self-dual vector potential case ($Q = \pm\, P$).

Both Eqs. (\ref{BS3}) and (\ref{FL3}) are expressed in the standard forms, which are our expected
consistent thermodynamical first law and Bekenstein-Smarr mass formula for the four-dimensional
dyonic KN-NUT spacetime. They are not only simple, but also unique, unlike the work \cite{JHEP0520084}
which declared that there are several different versions for them.

%%%%%%%%%%%%%%%%%%%%%%%%%%%%%%%%%%%%%%%%
\section{Reduced mass formulas}\label{V}
%%%%%%%%%%%%%%%%%%%%%%%%%%%%%%%%%%%%%%%%

In the subsections (\ref{II.A}, \ref{III.A}, \ref{IV.A}), the secondary hair: $J_N = MN$ has been
viewed as a independent thermodynamic variable, its impact on the thermodynamical relations has been
ignored. In this section, we will investigate this issue and derive the corresponding reduced mass
formulas of the dyonic RN-NUT, dyonic RN-NUT-AdS$_4$ and dyonic KN-NUT spacetimes, respectively. This
is somewhat analogous to those about the ``chirality condition": $J = Ml$ ($l$ is the cosmological
radius) of the superentropic Kerr-Newman-AdS$_4$, ultraspinning Kerr-Sen-AdS$_4$ and ultraspinning
dyonic Kerr-Sen-AdS$_4$ black holes \cite{PRD103-044014,PRD101-024057,PRD102-044007}.

Now considering $J_N = MN$ as a redundant variable and taking into account its differentiation $dJ_N
= M\, dN +N\, dM$, followed by eliminating them from the differential and integral mass formulas with
the help of $N = J_N/M$, then the first law and Bekenstein-Smarr mass formula boil down to their
nonstandard forms, which are listed below for the spacetimes considered before.

\textbf{II. A)} Dyonic RN-NUT spacetime:
\bea
(1 -\omega_h\, N)dM &=& (\kappa/2) d\cA_h +\bar{\psi}_h\, dN +\hat{\Phi}dQ +\hat{\Psi}dP \, , \nn \\
(1 -\omega_h\, N)M &=& \kappa\cA_h +\bar{\psi}_h\, N +\hat{\Phi}Q +\hat{\Psi}P \, ; \nn
\eea

\textbf{III. A)} Dyonic RN-NUT-AdS$_4$ spacetime:
\bea
(1 -\omega_h\, N)dM &=& (\kappa/2)d\cA_h +\bar{\psi}_h\, dN +\hat{\Phi}dQ +\hat{\Psi}dP \nn \\
&& +\cV d\cP \, , \nn \\
(1 -\omega_h\, N)M &=& \kappa\cA_h +\bar{\psi}_h\, N +\hat{\Phi}Q +\hat{\Psi}P -2\cV\cP \, ; \nn
\eea

\textbf{IV. A)} Dyonic KN-NUT spacetime:
\bea
(1 -\omega_h\, N)dM &=& (\kappa/2) d\cA_h +\Omega\, dJ +\bar{\psi}_h\, dN \nn \\
&& +\hat{\Phi}dQ +\hat{\Psi}dP \, , \nn \\
(1 -\omega_h\, N)M &=& \kappa\cA_h +2\Omega\, J +\bar{\psi}_h\, N +\hat{\Phi}Q +\hat{\Psi}P \, , \nn
\eea
where $\bar{\psi}_h = \psi_h +\omega_h\, M$ in each case.

It is easy to see that all of the thermodynamic quantities in these reduced mass formulas cannot
constitute the ordinary canonical conjugate pairs due to the presence of a factor $(1 -\omega_h\, N)$
in front of $dM$ and $M$. By the way, we mention that the above nonstandard mass formulas partially
appeared in some papers \cite{ApP22-227,IJTP52-2802,MPLA30-1550170,EPL115-30003}.

%%%%%%%%%%%%%%%%%%%%%%%%%%%%%%%%%%%%%%
\section{Concluding remarks}\label{VI}
%%%%%%%%%%%%%%%%%%%%%%%%%%%%%%%%%%%%%%

In this paper, we have extended our previous work \cite{PRD100-101501} to the more general
four-dimensional dyonic NUT-charged cases and followed a simple, systematic way to naturally
derive the thermodynamical first law and Bekenstein-Smarr mass formula via differentiating the
Christodoulou-Ruffini-like squared-mass formula with respect to its thermodynamic variables. If
only a secondary hair: $J_N = MN$ is included as did before, then the obtained thermodynamical
conjugate pairs fulfill the standard forms of the differential and integral mass formulas, except
that the derived electrostatic and magneto-static potentials are not equal to those calculated by
the standard method. Then, we demonstrated that this contradiction can be rectified via further
introducing two new additional secondary hairs: $Q_N = QN$ and $P_N = PN$, together with their
thermodynamical conjugate potentials ($\Phi_N, \Psi_N$). We have determined some special cases
with no need to include them, of which the $Q^2 = P^2$ case with the self-dual and anti-self-dual
Abelian vector potentials is possible to be particularly interesting. After that, the impact of
the secondary hair: $J_N = MN$ on the thermodynamics and the reduced mass formulas are discussed.

Our work demonstrated that, not only can the beautiful Bekenstein-Hawking one-quarter area-entropy
relation be naturally restored, but also four laws of the usual black hole thermodynamics are completely
applicable to the dyonic Taub-NUT-type spacetimes. We are believed that the strategy proposed in our
papers has provided the best and simplest scheme to formulate the consistent thermodynamical relations
for the NUT-charged spacetimes. A most related issue is to investigate the consistent thermodynamics
of the four-dimensional NUT-charged spacetimes in the K-K theory \cite{PRD103-064045,PRD77-124022,
CQG28-032001} and the EMDA theory \cite{2111.06111}. We hope to report the related progress soon.

\begin{acknowledgments}
We are greatly indebted to the anonymous referee for his/her invaluable comments and good suggestions
to improve the presentations of this work. This work is supported by the National Natural Science
Foundation of China (NSFC) under Grant No. 11675130, and by the Doctoral Research Initiation Project
of China West Normal University under Grant No. 21E028.
\end{acknowledgments}

\appendix

\section*{Appendix}

Currently, there exist three different fashions to formulate the `consistent' first law of the
four-dimensional NUT-charged spacetimes: I) Keeping the thermodynamic mass unmodified and introducing
new global-like charges (secondary hairs) and their conjugate potentials \cite{PRD100-101501}; II)
Retaining the thermodynamic mass unchanged and introducing new non-global Misner charge and its conjugate
conjugate variable \cite{PRD100-064055,PLB798-134972,JHEP0719119,CQG36-194001,JHEP0520084,PRD100-104016};
III) Only modifying the thermodynamic mass by including the contribution from the Misner string
\cite{PRD101-124011}. In our formalism \cite{PRD100-101501}, the consistent mass formulas are unique,
and every expressions for the thermodynamical quantities are very simple and concise. By contrast, not
only can the `consistent' first law of the NUTty dyonic spacetimes have ironically different possibilities
to be formulated as the electric-type, magnetic-type, mixed-type versions \cite{JHEP0719119,JHEP0520084},
and even many others \cite{PRD100-104016}, but also the expressions of some related thermodynamical
quantities are very complicated. Below, we will show that there are infinitely many `consistent' mass
formulas for the dyonic RN-NUT-AdS$_4$ spacetime if the thermodynamic mass is modified \emph{a la} the
mode proposed by Awad et al. \cite{PRD101-124011}.

%%%%%%%%%%%%%%%%%%%%%%%%%%%%%%%%%%%%%%%%%%%%%%%%%%%
\section{Infinitely many `consistent' mass formulas
for dyonic RN-NUT-AdS$_4$ spacetime}\label{appA}
%%%%%%%%%%%%%%%%%%%%%%%%%%%%%%%%%%%%%%%%%%%%%%%%%%%

In the spirit of Ref. \cite{PRD101-124011}, we consider to modify the thermodynamic mass which
receives the contribution from the Misner string so that the first law retains its usual form
without introducing new thermodynamical conjugate pairs. Then we can find that there are infinitely
many `consistent' mass formulas for the dyonic RN-NUT-AdS$_4$ spacetime as follows:
\bea
\widetilde{M} &=& 2T S +\Phi\widetilde{Q} +\Psi\widetilde{P} -2\widetilde{\cV}\mathcal{P} +N\chi \, , \\
d\widetilde{M} &=& T dS +\Phi\, d\widetilde{Q} +\Psi\, d\widetilde{P}
 +\widetilde{\cV}\, d\mathcal{P} +\chi\, dN \, ,
\eea
where the expressions of ($T, S, \Phi, \Psi)$ are given in Sec. \ref{III}, the pressure is
$\mathcal{P} = 3g^2/(8\pi)$, and
\bea
&& \widetilde{M} = M -N\chi \, , \quad \widetilde{\cV} = \frac{4\pi}{3}r_h(r_h^2 +3N^2) \, , \nn \\
&& \widetilde{Q} = Q +(w-1)N\Psi -2w_1N\Phi\, , \nn \\
&& \widetilde{P} = P +(w+1)N\Phi -2w_2N\Psi\, , \nn
\eea
together with a new thermodynamic potential conjugate to the NUT charge $N$:
\bea
\chi &=& N\frac{M -r_h+g^2r_h(r_h^2-3N^2)}{r_h^2+N^2} \nn \\
&& -w\Phi\Psi +w_1\Phi^2 +w_2\Psi^2 \nn \\
&=& \frac{-N}{2r_h}\big[1 -3g^2(r_h^2-N^2)\big]
 +\frac{N(Q^2+P^2)}{2r_h(r_h^2+N^2)} \nn \\
&& -w\Phi\Psi +w_1\Phi^2 +w_2\Psi^2 \, , \nn
\eea
in which $w, w_1, w_2$ are three arbitrary constants.

Some special cases may be very interesting: Let $w_1 = w_2 = 0$, then $\widetilde{Q} = Q$ and
$\widetilde{P} = P +2N\Phi\equiv\, P_h$ when $w = 1$; whilst $\widetilde{Q} = Q -2N\Psi\equiv\, Q_h$
and $\widetilde{P} = P$ when $w = -1$. A most simple case is to set $w = w_1 = w_2 = 0$, then $\chi$
is proportional to the value of the NUT (twist) potential measured at the horizon relative to the
infinity. Since $w, w_1, w_2$ can take arbitrary values, then a natural consequence is that there
are many different versions for the Lorentzian dyonic RN-NUT-AdS$_4$ spacetime.

\bigskip
%%%%%%%%%%%%%%%%%%%%%%%%%%%%%%%%%%%%%%%%%%%%%%%%%%%%%%%%%%%%%%
\section{Another type of mass formulas for Kerr-NUT spacetime}
%%%%%%%%%%%%%%%%%%%%%%%%%%%%%%%%%%%%%%%%%%%%%%%%%%%%%%%%%%%%%%

Extend to the rotating Kerr-NUT case, another `consistent' mass formulas are given below:
\bea
\widetilde{M} &=& 2T S +2\Omega\widetilde{J} +N\chi \, , \\
d\widetilde{M} &=& T dS +\Omega\, d\widetilde{J} +\chi\, dN \, ,
\eea
where
\bea
&& T = \frac{r_h -M}{2\pi\cA_h} \, , \quad \Omega = \frac{a}{\cA_h} \, , \quad
\chi = \frac{-N}{2r_h}\nn \\
&& \widetilde{M} = M -N\chi \, , \quad \widetilde{J} = (M -2N\chi)a \, , \quad
 S = \pi\cA_h \, . \nn
\eea
in which $\cA_h = r_h^2 +a^2 +N^2 = 2Mr_h +2N^2$.

\bigskip
Clearly, the above two appendices show that there are many different versions of the `consistent'
first law and Bekenstein-Smarr mass formula in the Lorentzian Taub-NUT-type spacetimes by modifying
the thermodynamic mass without introducing a new thermodynamical conjugate pair. Thus, a natural
puzzle is: Which version of the `consistent' first law is the most appropriate one?

% \bibliography{}

\begin{thebibliography}{99}
%\def\CQG{Class. Quant. Grav.\,}
\def\CQG{Classical Quantum Gravity\,}
\def\GRG{Gen. Relativ. Gravit.\,}
%\def\GRG{Gen. Rel. Grav.\,}
\def\JHEP{J. High Energy Phys.\,}
\def\PRD{Phys. Rev. D\,}
\def\PRL{Phys. Rev. Lett.\,}
\def\NPB{Nucl. Phys. \,}
\def\PLB{Phys. Lett. B \,}
\def\JMP{J. Math. Phys. (N.Y.)\,}
\def\AP{Ann. Phys. (N.Y.)\,}
\def\CPL{Chin. Phy. Lett.\,}

\bibitem{PRD100-064055}
R.A. Hennigar, D. Kubiz\v{n}\'ak, and R.B. Mann,
\textit{Thermodynamics of Lorentzian Taub-NUT spacetimes},
\href{http://dx.doi.org/10.1103/PhysRevD.100.064055}
{\PRD \textbf{100}, 064055 (2019)}.
%[\href{http://arxiv.org/abs/1903.8668}{arXiv:1903.08668}].

\bibitem{PLB798-134972}
A.B. Bordo, F. Gray, R.A. Hennigar, and D. Kubiz\v{n}\'ak,
\textit{The first law for rotating NUTs},
\href{https://doi.org/10.1016/j.physletb.2019.134972}
{\PLB \textbf{798}, 134972 (2019)}.
%[\href{http://arxiv.org/abs/1905.06350}{arXiv:1905.06350}].

\bibitem{JHEP0719119}
A.B. Ballon, F. Gray, and D. Kubiz\v{n}\'ak,
\textit{Thermodynamics and phase transitions of NUTty dyons},
\href{https://doi.org/10.1007/JHEP07(2019)119}
{\JHEP \textbf{07} (2019) 119}.
%[\href{https://arxiv.org/abs/1904.00030}{arXiv:1904.00030}].

\bibitem{CQG36-194001}
A.B. Bordo, F. Gray, R.A. Hennigar, and D. Kubiz\v{n}\'ak,
\textit{Misner gravitational charges and variable string strengths},
\href{https://doi.org/10.1088/1361-6382/ab3d4d}
{\CQG \textbf{36}, 194001 (2019)}.
%[\href{https://arxiv.org/abs/arXiv:1905.03785}{arXiv:1905.03785}].

\bibitem{JHEP0520084}
A.B. Ballon, F. Gray, and D. Kubiz\v{n}\'ak,
\textit{Thermodynamics of rotating NUTty dyons},
\href{https://doi.org/10.1007/JHEP05(2020)084}
{\JHEP \textbf{05} (2020) 084}.
%[\href{https://arxiv.org/abs/2003.02268}{arXiv:2003.02268}].

\bibitem{PLB802-135270}
G. Cl\'ement and D. Gal'tsov,
\textit{On the Smarr formulas for electrovac spacetimes with line singularities},
\href{https://doi.org/10.1016/j.physletb.2020.135270}
{\PLB \textbf{802}, 135270 (2020)}.
%[\href{http://arxiv.org/abs/1908.10617}{arXiv:1908.10617}].

\bibitem{1908.04238}
R. Durka,
\textit{The first law of black hole thermodynamics for Taub-NUT spacetime},
\href{http://arxiv.org/abs/1908.04238}{arXiv:1908.04238}.

\bibitem{PRD100-104016}
Z.H. Chen and J. Jiang,
\textit{General Smarr relation and first law of a NUT dyonic black hole},
\href{http://dx.doi.org/10.1103/PhysRevD.100.104016}
{\PRD \textbf{100}, 104016 (2019)}.
%[\href{http://arxiv.org/abs/1910.10107}{arXiv:1910.10107}].

\bibitem{JHEP0821152}
N. Abbasvandi, M. Tavakoli, and R.B. Mann,
\textit{Thermodynamics of dyonic NUT charged black holes with entropy as Noether charge},
\href{https://doi.org/10.1007/JHEP08(2021)152}
{\JHEP \textbf{08} (2021) 152}.
%[\href{http://arxiv.org/abs/2107.00182}{arXiv:2107.00182}].

\bibitem{2109.07715}
E. Frodden and D. Hidalgo,
\textit{The first law for the Lorentzian rotating Taub-NUT},
\href{http://arxiv.org/abs/2109.07715}{arXiv:2109.07715}.

\bibitem{2112.00780}
N.H. Rodriguez and M.J. Rodriguez,
\textit{First law for Kerr Taub-NUT AdS black holes},
\href{http://arxiv.org/abs/2112.00780}{arXiv:2112.00780}.

\bibitem{PRD101-124011}
A.M. Awad and S. Eissa,
\textit{Topological dyonic Taub-Bolt/NUT-AdS solutions: Thermodynamics and first law},
\href{http://dx.doi.org/10.1103/PhysRevD.101.124011}
{\PRD \textbf{101}, 124011 (2020)}.
%[\href{http://arxiv.org/abs/2007.10489}{arXiv:2007.10489}].

\bibitem{PRD100-101501}
S.-Q. Wu and D. Wu,
\textit{Thermodynamical hairs of the four-dimensional Taub-Newman-Unti-Tamburino spacetimes},
\href{http://dx.doi.org/10.1103/PhysRevD.100.101501}
{\PRD \textbf{100}, 101501(R) (2019)}.
%[\href{http://arxiv.org/abs/1909.07776}{arXiv:1909.07776}].
[In the Bekenstein-Smarr (or integral) mass formula of this reference, $J_n$ should be $J_N$
in three equations from (24) to (31); There is also a factor $r_h$ missed in the expression of
$\widetilde{\cV}$ which should be $4\pi\,r_h(r_h^2+3N^2)/3$ in the last 11-th line of the right
column on Page 5.]

\bibitem{PRL25-1596}
D. Christodoulou,
\textit{Reversible and Irreversible Transforations in Black Hole Physics},
\href{http://dx.doi.org/10.1103/PhysRevLett.25.1596}
{\PRL \textbf{25}, 1596 (1970)}.

\bibitem{PRD4-3552}
D. Christodoulou and R. Ruffini,
\textit{Reversible transformations of a charged black hole},
\href{http://dx.doi.org/10.1103/PhysRevD.4.3552}
{\PRD \textbf{4}, 3552 (1971)}.

\bibitem{PLB634-531}
R.B. Mann and C. Stelea,
\textit{On the gravitational energy of the Kaluza Klein monopole},
\href{https://doi.org/10.1016/j.physletb.2006.02.025}
{\PLB \textbf{634}, 531 (2006)}.
%[\href{https://arxiv.org/abs/hep-th/0511180}{hep-th/0511180}]

\bibitem{PRD77-044038}
A.N. Aliev,
\textit{Rotating spacetimes with asymptotic nonflat structure and the gyromagnetic ratio},
\href{http://dx.doi.org/10.1103/PhysRevD.77.044038}
{\PRD \textbf{77}, 044038 (2008)}.
%[\href{http://arxiv.org/abs/arXiv:0711.4614}{arXiv:0711.4614}].

\bibitem{CQG3-65}
M. Mueller and M.J. Perry,
\textit{Constraints on magnetic mass},
\href{https://doi.org/10.1088/0264-9381/3/1/009}
{\CQG \textbf{3}, 65 (1986)}.

\bibitem{PPS92-1}
J.S. Dowker and J.A. Roche,
\textit{The gravitational analogues of magnetic monopoles},
\href{https://doi.org/10.1088/0370-1328/92/1/302}
{Proc. Phys. Soc. \textbf{92}, 1 (1967)}.

\bibitem{GRG5-603}
J.S. Dowker,
\textit{The nut solution as a gravitational dyon},
\href{https://doi.org/10.1007/BF02451402}
{\GRG \textbf{5}, 603 (1974)}.

\bibitem{PLB807-135521}
P. Pradhan,
\textit{Area (or entropy) products for Newman-Unti-Tamburino class of black holes},
\href{https://doi.org/10.1016/j.physletb.2020.135521}
{\PLB \textbf{807}, 135521 (2020)}.
%[\href{http://arxiv.org/abs/arXiv:206.15092}{arXiv:2006.15092}].

\bibitem{PRL106-121301}
M. Cveti\v{c}, G.W. Gibbons, and C.N. Pope,
\textit{Universal Area Product Formulae for Rotating and Charged Black Holes in Four and Higher Dimensions},
\href{http://dx.doi.org/10.1103/PhysRevLett.106.121301}
{\PRL \textbf{106}, 121301 (2011)}.
%[\href{https://arxiv.org/abs/1011.0008}{arXiv:1011.0008}].

\bibitem{GRG53-69}
P. Pradhan,
\textit{Energy formula for Newman-Unti-Tamburino class of black holes},
\href{https://doi.org/10.1007/s10714-021-02836-w}
{\GRG \textbf{53}, 69 (2021)}.
%[\href{http://arxiv.org/abs/arXiv:2107.09578}{arXiv:2107.09578}].

\bibitem{CQG19-2051}
A.M. Awad and A. Chamblin,
\textit{A bestiary of higher dimensional Taub-NUT AdS space-times},
\href{https://doi.org/10.1088/0264-9381/19/8/301}
{\CQG \textbf{19}, 2051 (2003)}.

\bibitem{CQG23-2849}
A.M. Awad,
\textit{Higher dimensional Taub-NUTS and Taub-Bolts in Einstein-Maxwell gravity},
\href{https://doi.org/10.1088/0264-9381/23/9/006}
{\CQG \textbf{23}, 2849 (2006)}.

\bibitem{CQG21-2937}
R.B. Mann and C. Stelea,
\textit{Nuttier (A)dS black holes in higher dimensions},
\href{https://doi.org/10.1088/0264-9381/21/12/010}
{\CQG \textbf{21}, 2937 (2004)}.

\bibitem{PLB634-448}
R.B. Mann and C. Stelea,
\textit{New multiply nutty spacetimes},
\href{https://doi.org/10.1016/j.physletb.2006.02.019}
{\PLB \textbf{634}, 448 (2006)}.

\bibitem{PLB632-537}
R.B. Mann and C. Stelea,
\textit{New Taub-NUT-Reissner-Nordstrom spaces in higher dimensions},
\href{https://doi.org/10.1016/j.physletb.2005.10.085}
{\PLB \textbf{632}, 537 (2006)}.

\bibitem{CQG23-5323}
W. Chen, H. L\"u, and C.N. Pope,
\textit{General Kerr-NUT-AdS metrics in all dimensions},
\href{https://doi.org/10.1088/0264-9381/23/17/013}
{\CQG \textbf{23}, 5323 (2006)}.

\bibitem{NPB762-38}
W. Chen, H. L\"u, and C.N. Pope,
\textit{Kerr-de Sitter black holes with NUT charges},
\href{https://doi.org/10.1016/j.nuclphysb.2006.07.022}
{\NPB \textbf{B762}, 38 (2007)}.

\bibitem{PRD103-064045}
I. Bogush, G. Cl\'ement, D. Gal'tsov, and D. Torbunov,
\textit{Nutty Kaluza-Klein dyons revisited},
\href{http://dx.doi.org/10.1103/PhysRevD.103.064045}
{\PRD \textbf{103}, 064045 (2021)}.

\bibitem{PRD77-124022}
A.N. Aliev, H. Cebeci, and T. Dereli,
\textit{Kerr-Taub-NUT spacetime with Maxwell and dilaton fields},
\href{http://dx.doi.org/10.1103/PhysRevD.77.124022}
{\PRD \textbf{77}, 124022 (2008)}.

\bibitem{CQG28-032001}
D.D.K. Chow,
\textit{Single-charge rotating black holes in four-dimensional gauged supergravity},
\href{https://doi.org/10.1088/0264-9381/23/17/013}
{\CQG \textbf{28}, 032001 (2011)}.

\bibitem{2111.06111}
D. Gal'tsov, G. Cl\'ement, and I. Bogush,
\textit{Einstein-Maxwell-Dilaton-Axion mass formulas for black holes with struts and strings},
\href{http://arxiv.org/abs/2111.06111}{arXiv:2111.06111}.

\bibitem{NPB717-246}
Z.-W. Chong, M. Cveti\v{c}, H. L\"u, and C.N. Pope,
\textit{Charged rotating black holes in four-dimensional gauged and ungauged supergravities},
\href{https://doi.org/10.1016/j.nuclphysb.2005.03.034}
{\NPB \textbf{B717}, 246 (2005)}.

\bibitem{CQG31-022001}
D.D.K. Chow and G. Comper\'e,
\textit{Seed for general rotating non-extremal black holes of $\cN = 8$ supergravity},
\href{https://doi.org/10.1088/0264-9381/31/2/022001}
{\CQG \textbf{31}, 022001 (2014)}.

\bibitem{PRD90-025029}
D.D.K. Chow and G. Comper\'e,
\textit{Black holes in $\cN = 8$ supergravity from SO(4,4) hidden symmetries},
\href{http://dx.doi.org/10.1103/PhysRevD.90.025029}
{\PRD \textbf{90}, 025029 (2014)}.

\bibitem{JMP4-915}
E.T. Newman, L. Tamburino, and T. Unti,
\textit{Empty space generalization of the Schwarzschild metric},
\href{https://doi.org/10.1063/1.1704018}
{\JMP \textbf{4}, 915 (1963)}.

\bibitem{AP98-98}
J.F. Plebanski and M. Demianski,
\textit{Rotating, charged, and uniformly accelerating mass in general relativity},
\href{https://doi.org/10.1016/0003-4916(76)90240-2}
{\AP \textbf{98}, 98 (1976)}.

\bibitem{JMP4-924}
C.W. Misner,
\textit{The flatter regions of Newman, Unti, and Tamburino's generalized Schwarzschild space},
\href{https://doi.org/10.1063/1.1704019}
{\JMP \textbf{4}, 924 (1963)}.

\bibitem{PR133-B845}
D.R. Brill,
\textit{Electromagnetic fields in a homogeneous, non-isotropic universe},
\href{http://dx.doi.org/10.1103/PhysRev.133.B845}
{Phys. Rev. \textbf{133}, B845 (1964)}.

\bibitem{CWM}
C.W. Misner,
\textit{Taub-NUT space as a counter-example to almost anything}, in J. Ehlers ed.,
Relativity Theory and Astrophysics I: Relativity and Cosmology, Lectures in Applied
Mathematics (American Mathematical Society, Providence, 1967), Vol. \textbf{8}, pp. 160.

\bibitem{PLB750-591}
G. Cl\'ement, D. Gal'tsov, and M. Guenouche,
\textit{Rehabilitating space-times with NUTs},
\href{https://doi.org/10.1016/j.physletb.2015.09.074}
{\PLB \textbf{750}, 591 (2015)}.
%[\href{http://arxiv.org/abs/1508.07622}{arXiv:1508.07622}].

\bibitem{PRD93-024048}
G. Cl\'ement, D. Gal'tsov, and M. Guenouche,
\textit{NUT wormholes},
\href{https://doi.org/10.1103/PhysRevD.93.024048}
{\PRD \textbf{93}, 024048 (2016)}.
%[\href{http://arxiv.org/abs/1509.07854}{arXiv:1509.07854}].

\bibitem{GRG50-60}
G. Cl\'ement and M. Guenouche,
\textit{Motion of charged particles in a NUTty Einstein-Maxwell spacetime and causality violation},
\href{https://doi.org/10.1007/s10714-018-2388-y}
{\GRG \textbf{50}, 60 (2018)}.
%[\href{http://arxiv.org/abs/1606.08457}{arXiv:1606.08457}].

\bibitem{PCPS66-145}
W.B. Bonnor,
\textit{A new interpretation of the NUT metric in general relativity},
\href{https://doi.org/10.1017/S0305004100044807}
{Proc. Cambridge Philos. Soc. \textbf{66}, 145 (1969)}.

\bibitem{PCPS70-89}
A. Sackfield,
\textit{Physical interpretation of NUT metric},
\href{https://doi.org/10.1017/S0305004100049707}
{Proc. Cambridge Philos. Soc. \textbf{70}, 89 (1971)}.

\bibitem{CQG22-3555}
V.S. Manko and E. Ruiz,
\textit{Physical interpretation of the NUT family of solutions},
\href{https://doi.org/10.1088/0264-9381/22/17/014}
{\CQG \textbf{22}, 3555 (2005)}.
%[\href{https://arxiv.org/abs/gr-qc/0505001}{gr-qc/0505001}]

\bibitem{CQG23-4473}
V.S. Manko, J. Martin, and E. Ruiz,
\textit{Singular sources in the Demianski-Newman spacetimes},
\href{https://doi.org/10.1088/0264-9381/23/13/011}
{\CQG \textbf{23}, 4473 (2006)}.
%[\href{http://arxiv.org/abs/gr-qc/0603002}{gr-qc/0603002}].

\bibitem{PLB608-251}
S.-Q. Wu,
\textit{New formulation of the first law of black hole thermodynamics: A stringy analogy},
\href{https://doi.org/10.1016/j.physletb.2005.01.018}
{\PLB \textbf{608}, 251 (2005)}.
%[\href{https://arxiv.org/abs/gr-qc/0405029}{gr-qc/0405029}]

\bibitem{JMP22-2612}
S. Ramaswamy and A. Sen,
\textit{Dual-mass in general relativity},
\href{https://doi.org/10.1063/1.524839}
{\JMP \textbf{22}, 2612 (1981)}.

\bibitem{JMP23-2168}
A. Ashtekar and A.Sen,
\textit{NUT 4-momenta are forever},
\href{https://doi.org/10.1063/1.525274}
{\JMP \textbf{23}, 2168 (1982)}.

\bibitem{DN1966}
M. Demianski and E.T. Newman,
\textit{A combined Kerr-NUT solution of Einstein field equation},
\href{http://adsabs.harvard.edu/abs/1966BAPSS...14.653N}
{Bull. Acad. Pol. Sci., Ser. Sci., Math., Astron. Phys. \textbf{14}, 653 (1966)}.

\bibitem{PRD59-024009}
C.J. Hunter,
\textit{The action of instantons with NUT charge},
\href{https://doi.org/10.1103/PhysRevD.59.024009}
{\PRD \textbf{59}, 024009 (1999)}.
%[\href{http://arxiv.org/abs/gr-qc/9807010}{gr-qc/9807010}].

\bibitem{PRD103-044014}
D. Wu, S.-Q. Wu, P. Wu, and H. Yu,
\textit{Aspects of the dyonic Kerr-Sen-AdS$_4$ black hole and its ultraspinning version},
\href{http://dx.doi.org/10.1103/PhysRevD.103.044014}
{\PRD \textbf{103}, 044014 (2021)}.
%[\href{http://arxiv.org/abs/arXiv:2010.13518}{arXiv:2010.13518}].

\bibitem{PRD101-024057}
D. Wu, P. Wu, H. Yu, and S.-Q. Wu,
\textit{Notes on the thermodynamics of superentropic AdS black holes},
\href{http://dx.doi.org/10.1103/PhysRevD.101.024057}
{\PRD \textbf{101}, 024057 (2020)}.
%[\href{https://arxiv.org/abs/1909.07776}{arXiv:1912.03576}].

\bibitem{PRD102-044007}
D. Wu, P. Wu, H. Yu, and S.-Q. Wu,
\textit{Are ultraspinning Kerr-Sen-AdS$_4$ black holes always superentropic?},
\href{http://dx.doi.org/10.1103/PhysRevD.102.044007}
{\PRD \textbf{102}, 044007 (2020)}.
%[\href{https://arxiv.org/abs/2007.02224}{arXiv:2007.02224}].

\bibitem{PRD21-884}
D.C. Wright,
\textit{Black holes and the Gibbs-Duhem relation},
\href{http://dx.doi.org/10.1103/PhysRevD.21.884}
{\PRD \textbf{21}, 884 (1980)}.

\bibitem{CQG17-399}
M.M. Caldarelli, G. Cognola, and D. Klemm,
\textit{Thermodynamics of Kerr-Newman-AdS black holes and conformal field theories},
\href{http://dx.doi.org/10.1088/0264-9381/17/2/310}
{\CQG \textbf{17}, 399 (2000)}.
%[\href{http://arxiv.org/abs/hep-th/9908022}{hep-th/9908022}].

\bibitem{CPL23-1096}
S. Wang, S.-Q. Wu, F. Xie, and L. Dan,
\textit{The first laws of thermodynamics of the (2+1)-dimensional BTZ black holes and Kerr-de Sitter
 spacetimes},
\href{http://dx.doi.org/10.1088/0256-307X/23/5/009}
{\CPL \textbf{23}, 1096 (2006)}.
%[\href{https://arxiv.org/abs/hep-th/0601147}{hep-th/0601147}].

\bibitem{PRD93-084022}
R. Araneda, R. Aros, O. Miskovic, and R. Olea,
\textit{Magnetic mass in 4d AdS gravity},
\href{https://doi.org/10.1103/PhysRevD.93.084022}
{\PRD \textbf{93}, 084022 (2016)}.
%[\href{https://arxiv.org/abs/1602.07975}{arXiv:1602.07975}].

\bibitem{CQG31-235003}
C.V. Johnson,
\textit{Thermodynamic volumes for AdS-Taub-NUT and AdS-Taub-Bolt},
\href{https://doi.org/10.1088/0264-9381/31/23/235003}
{\CQG \textbf{31}, 235003 (2014)}.
%[\href{https://arxiv.org/abs/arXiv:1405.5941}{arXiv:1405.5941}].

\bibitem{ApP22-227}
M.H. Ali,
\textit{Planck absolute entropy of Demianski-Newman black holes},
\href{https://doi.org/10.1016/j.astropartphys.2004.06.002}
{Astropart. Phys. \textbf{22}, 227 (2004)}.

\bibitem{IJTP52-2802}
M.H. Ali and K. Sultana,
\textit{Charged particles' Hawking radiation via tunneling of both horizons from
Reissner-Nordstrom-Taub-NUT black holes},
\href{https://doi.org/10.1007/s10773-013-1572-9}
{Int. J. Theor. Phys. \textbf{52}, 2802 (2013)}.

\bibitem{MPLA30-1550170}
P. Pradhan,
\textit{Area product and mass formula for Kerr-Newman-Taub-NUT spacetime},
\href{https://doi.org/10.1142/S0217732315501709}
{Mod. Phys. Lett. A \textbf{30}, 1550170 (2015)}.
%[\href{https://arxiv.org/abs/1310.7921}{arXiv:1310.7921}].

\bibitem{EPL115-30003}
P. Pradhan,
\textit{Surface area products for Kerr-Taub-NUT space-time},
\href{https://doi.org/10.1209/0295-5075/115/30003}
{Europhys. Lett. \textbf{115}, 30003 (2016)}.
%[\href{https://arxiv.org/abs/1408.2973}{arXiv:1408.2973}].

\end{thebibliography}

\end{document}